\documentclass[]{spie}  

\usepackage{amsmath,amsfonts,amssymb}
\usepackage{graphicx}
\usepackage{tabularx}
\usepackage[colorlinks=true, allcolors=blue]{hyperref}
\usepackage{wrapfig}
\usepackage{url}

\title{Towards Augmented Reality-based Suturing in \\ Monocular Laparoscopic Training}

\author[a,b]{Chandrakanth Jayachandran Preetha}
\author[a,b]{Jonathan Kloss}
\author[a]{Fabian Siegfried Wehrtmann}
\author[c]{Lalith Sharan}
\author[a]{Carolyn Fan}
\author[a]{Beat Peter M\"{u}ller-Stich}
\author[a]{Felix Nickel}
\author[c]{Sandy~Engelhardt}

\affil[a]{Department of General, Visceral, and Transplantation Surgery,
University Hospital Heidelberg, Germany}
\affil[b]{Otto-von-Guericke University Magdeburg, Germany}
\affil[c]{Working Group Artificial Intelligence in Cardiovascular Medicine,
University Hospital Heidelberg, Germany}

\begin{document} 
\maketitle
\begin{abstract}
Minimally Invasive Surgery (MIS) techniques have gained rapid popularity among surgeons since they offer significant clinical benefits including reduced recovery time and diminished post-operative adverse effects. However, conventional endoscopic systems output monocular video which compromises depth perception, spatial orientation and field of view. Suturing is one of the most complex tasks performed under these circumstances. Key components of this tasks are the interplay between needle holder and the surgical needle. Reliable 3D localization of needle and instruments in real time could be used to augment the scene with additional parameters that describe their quantitative geometric relation, e.g.\ the relation between the estimated needle plane and its rotation center and the instrument. This could contribute towards standardization and training of basic skills and operative techniques, enhance overall surgical performance, and reduce the risk of complications. 
The paper proposes an Augmented Reality environment with quantitative and qualitative visual representations to enhance laparoscopic training outcomes performed on a silicone pad. This is enabled by a multi-task supervised deep neural network which performs multi-class segmentation and depth map prediction. Scarcity of labels has been conquered by creating a virtual environment which resembles the surgical training scenario to generate dense depth maps and segmentation maps.  
The proposed convolutional neural network was tested on real surgical training scenarios and showed to be robust to occlusion of the needle. The network achieves a dice score of 0.67 for surgical needle segmentation, 0.81 for needle holder instrument segmentation and a mean absolute error of 6.5~mm for depth estimation.
\end{abstract}
\keywords{Quantitative Surgical Training, Augmented Reality,  Monocular Laparoscopy, Suturing, Segmentation, Depth Prediction}

\section{\hspace{14pt}Description of Purpose}

\begin{figure*}[!b]
   \begin{center}
 \includegraphics[width=\linewidth]{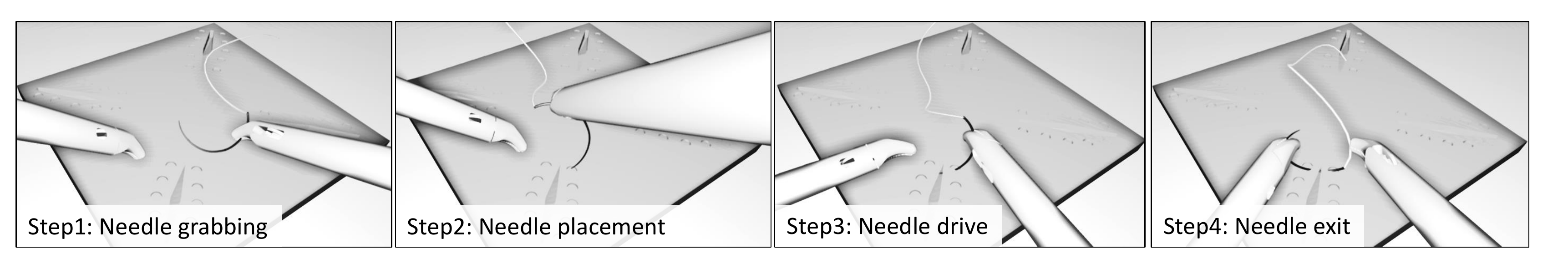}
   \end{center}
   \caption[] 
   { \label{fig:Steps} Four steps of surgical suturing.}
\end{figure*}

\label{paragraph:Motivation}
Minimally Invasive Surgery (MIS) techniques became popular in the last decades since they offer significant clinical benefits including reduced recovery time and fewer scars for the patients. However, most endoscopic systems rely on monocular video which is known to compromise depth perception, spatial orientation and field of view, making it more difficult for the surgeon to reliably perform suturing. 

In MIS, manual interaction between needle holder instrument and the surgical needle is a key requirement which needs to be constantly trained and exercised. However, active support and enhanced quantitative analysis for guidance of the suturing process is relatively unexplored. So far, suturing has been roughly described in many textbooks and tutorials by geometric relations between needle and instruments \cite{Liceaga,Hudgens}
, but live-support during surgical training or surgery itself is not available.

\begin{figure*}[t]
   \begin{center}
 \includegraphics[width=0.7\linewidth]{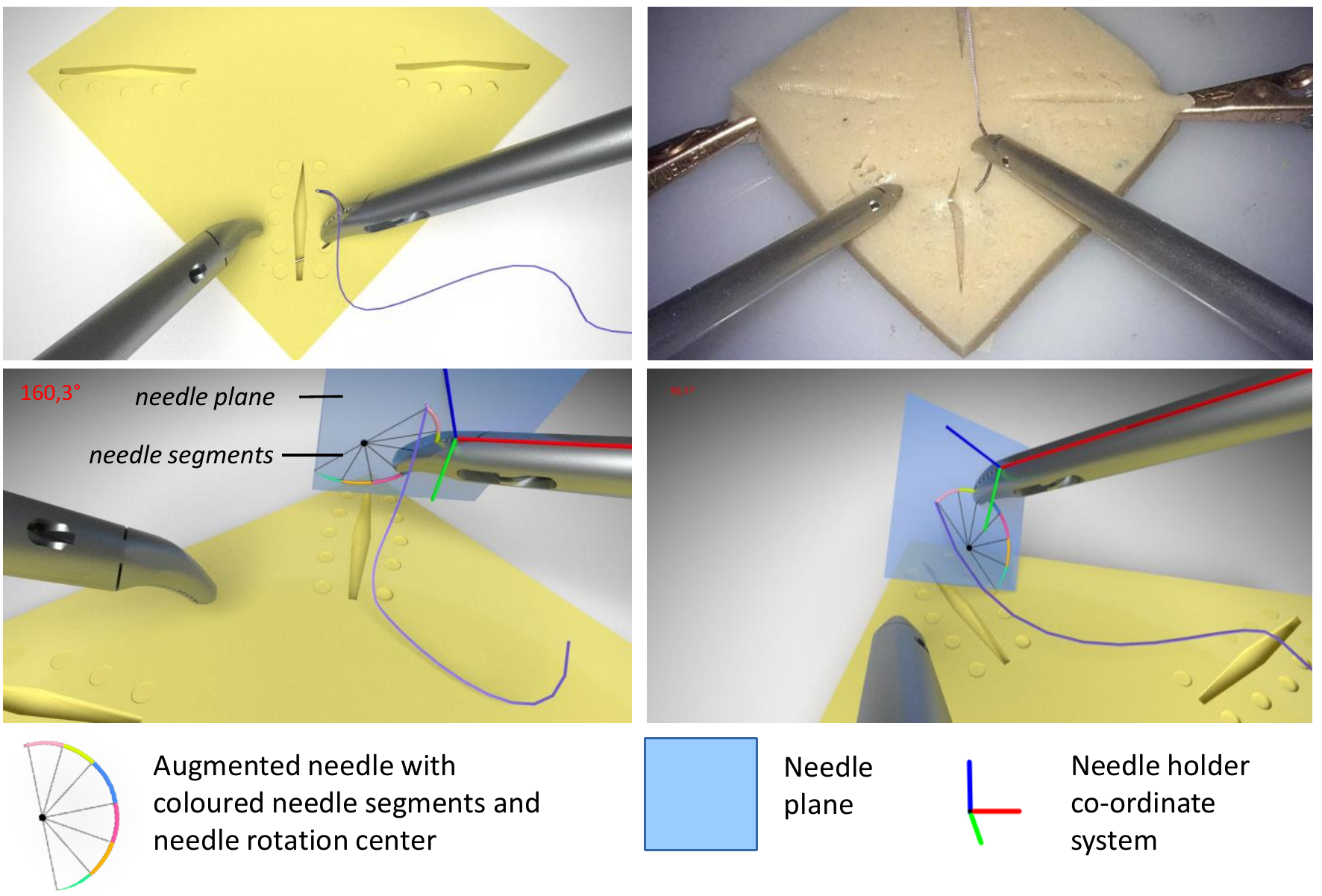}
   \end{center}
   \caption[] 
   { \label{fig:Examples} Top row: Typical training scenario recreated in virtual environment vs.\ real example. Bottom row: Mock-up of an augmented reality aided training environment to illustrate the novel use-case.}
\end{figure*}

Four relevant phases of the suturing process can be defined (cf. Fig. \ref{fig:Steps}):
\begin{enumerate}
\item{\textbf{Grasping the needle with the needle holder}:
 Position the needle to the needle holder at approximately 2/3 from the tip of the needle.
 Fix the needle at an angle between 90-120$^\circ$ regarding the needle holder`s axis. }
\item{\textbf{Placing the needle}:
 Place the tip of the needle towards the entry point of the tissue, forming a 90$^\circ$ entry angle of the tip of the needle with the tissue. }
\item{\textbf{Moving the hand}: 
The needle holder is rotated around its longitudinal axis to pass the needle through the tissue. Parallel planes between needle holder and wound should be maintained.}
\item{\textbf{Exiting the needle}:
Grasp the exiting needle with the second instrument at approximately 1/3 from the tip. Rotate it out of the tissue.}
\end{enumerate}
Common technical mistakes are a poorly positioned needle or a traumatic and inefficient handling of the needle and tissue (e.g.\ not performing pure rotation of the needle around its rotation center, but also pulling). This may lead to a considerable waste of time, and in general the whole process loses quality and efficiency.
Training labs offer possibilities for medical students or surgical residents to learn suturing on a laparoscopic box trainer to avoid such mistakes. 
However, the learning process is still not very effective and must be guided by an expert surgeon. At the same time, reliable 3D localization of needle and instruments in real time could be leveraged to augment the laparoscopic scene with additional \textit{quantitative parameters} and \textit{visual cues} that better describe their relation. We therefore propose a novel use-case for a laparoscopic Augmented Reality (AR) system as shown in Fig. \ref{fig:Examples}, which has not been described before.

Several sophisticated technical steps are necessary that perform segmentation, 3D scene understanding, tracking to realize such a system. 
While some of these steps, like segmentation and vision-based tracking of surgical instruments have been addressed by several works \cite{DBLP:journals/corr/abs-1805-02475,Bouget2017VisionbasedAM}
, and challenges (e.g. EndoVis\footnote{https://endovissub-instrument.grand-challenge.org/}), the same for the needle is relatively unexplored. 
In comparison to the instruments, the needle is considerably smaller and thinner, has shinier reflections and is most of the time partially occluded by the needle holder and tissue. Therefore, segmentation and tracking of the needle is a huge challenge. Speidel et al. \cite{Speidel2015ImagebasedTO} implemented a markerless needle tracking method which combines colour and geometry based approaches. The method, however, failed to produce robust segmentation of the needle in the presence of specular highlights, varying light conditions and occlusions. 

In this work, we solve first steps towards the vision of quantitative suturing support in surgical training. 
We propose a multi-task supervised convolutional neural network that is able to segment the needle  and the needle holders and that makes a depth estimation from a \textit{monocular} camera of the scene. In order to overcome the scarcity of annotations, we propose to create a virtual representation of a surgical training scenario that includes a suture pad as commonly used in our clinic to train medical students \cite{Schmidt2019SelfdirectedTW}. 
Furthermore, we propose example AR-visualizations that can be connected to  the recovered 3D information to guide the trainee.

\section{\hspace{14pt}Material and Methods}

Supervised training of neural networks for computer vision tasks requires large amounts of training data. Currently, manual annotations or additional sensor data are commonly used as ground truth for visual recognition tasks. However, manual annotation for low-level tasks like semantic segmentation is time-consuming and error-prone and dense depth information cannot be obtained by manual annotations at all. Beyond that, using additional laser sensors for objects very close to the laparoscope is difficult. 
Creating virtual environments for generation of synthetic data as input for learning tasks offers a viable solution to this problem, especially for a monocular approach that can not infer disparity by a second view. This has been investigated by other works in computer vision that try to solve e.g.\ semantic segmentation, optical flow or disparity estimation \cite{MIFDB18,7780721,Bhoi2019MonocularDE}. 

\subsection{Creation of a virtual environment}


A virtual environment consisting of 3D mesh models of needle holders, needle, thread and a silicone suture pad was created to closely match the arrangement of these objects in a laparoscopic box trainer. A virtual model of the CV-25, 1/2 circle needle was carefully designed.
An example scene is shown in Fig.\ \ref{fig:Examples} (upper row). The different material and illumination properties like colour, reflections and roughness of various objects were fine-tuned in order to produce renderings more closer to photo-realism. Careful attention was also paid to the generation of the 3D models so as to faithfully recreate the geometric properties of the original objects. In this work, we make use of Blender 2.79, a 3D modelling and animation software to create the virtual environment for the generation of synthetic images. These synthetic images form part of the training data (segmentation labels and depth information) for supervised training of the deep learning approach presented below. 



\subsection{Multi-task deep learning network}


\begin{figure*}[b]
   \begin{center}
 \includegraphics[width=\linewidth]{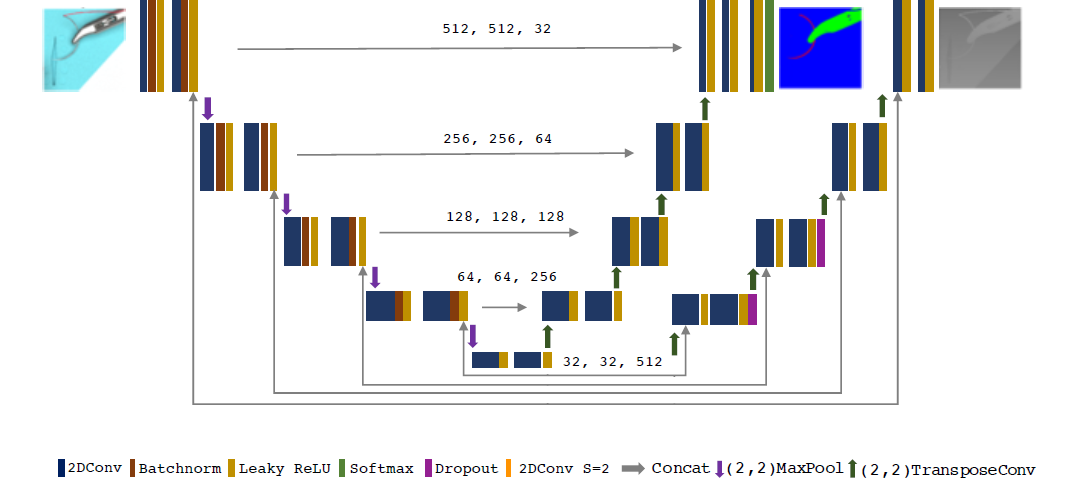}
   \end{center}
   \caption[] 
   { \label{fig:arch} Network architecture.}
\end{figure*}

For the purpose of solving segmentation and depth estimation, we developed a multi-task encoder-decoder architecture as illustrated in Fig.\ \ref{fig:arch}. 
The basic structure of the network relies on the U-Net \cite{DBLP:journals/corr/RonnebergerFB15} with two different decoders being used to predict the segmentation map and the depth map. 
Each stage of the encoder consists of two $(3,3)$ convolution layers followed by batch normalization and Leaky ReLU activation function. The $(2,2)$ max-pooling layers down-sample the feature maps at each level in the encoder. The model includes two decoders which specialize in depth map estimation and multi-class segmentation. The basic block in the decoder consists of two convolution layers followed by Leaky ReLU activation layer. The use of transposed convolution layers allows the network to learn optimal up-sampling of the feature maps and the skip connections to the decoder restore spatial information lost during down-sampling.




The multi-task learning strategy based on hard-parameter sharing of the encoder section acts as a regularization method and reduces the risk of overfitting \cite{DBLP:journals/corr/Ruder17a}. The network uses categorical cross-entropy and mean squared error (MSE) loss functions for multi-class segmentation and depth map estimation respectively.  The two branches of the network were initially trained simultaneously on the synthetic data set and the segmentation branch was subsequently retrained on real  images with annotations to improve the network's segmentation performance on real data. Shared encoder weights for segmentation and depth prediction enables the networks to jointly use this high-level information. 

\subsection{Augmented Reality Visualizations}

The proposed AR environment could assist the surgeon in the suturing task through additional visualization of 1) coloured needle segments with optimal center of rotation, 2) the plane of the needle and fixed coordinate system of the needle holder. 
These visualizations along with measurements of angle between the plane of the needle and the coordinate system of the instrument could enable the surgeons to grasp the needle at the recommended position and to maintain the correct  trajectory of the needle.

\subsection{Training Data}

\begin{figure*}[t]
   \begin{center}
 \includegraphics[width=0.8\linewidth]{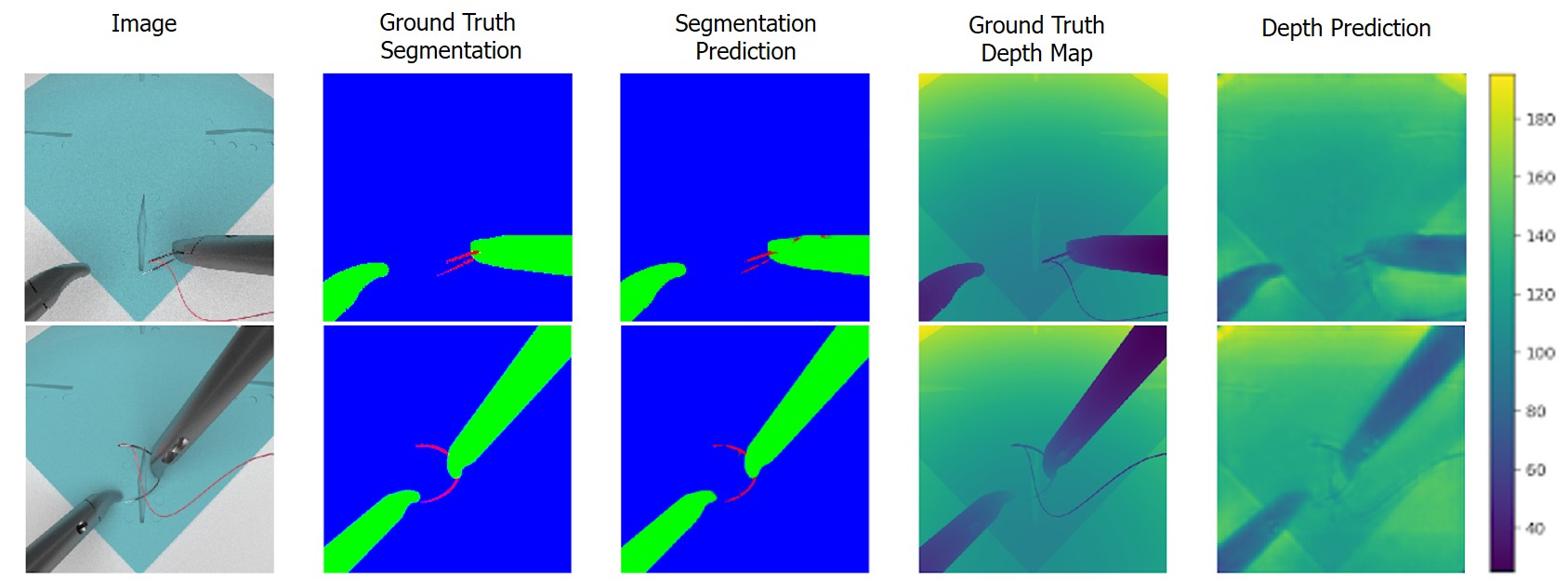}
   \end{center}
   \caption[] 
   { \label{fig:Results} Examples of network prediction for segmentation and depth estimation on synthetic test data. }
\end{figure*}

In total 218 frames were rendered in the virtual scene from different perspectives, including  random noise. 21 images were used for testing and on the 197 remaining images, data augmentation was performed. 
Images were reseized to 512 pixel along the $y$-axis and random crop of 512 pixel was performed along the $x$-axis. This resulted in 732 synthetic images used for training. 
In addition to the synthetic data set, video data from real laparoscopic training sequences was captured, segmented by a physician and randomly cropped, yielding 144 frames for training.
The model was first trained on a synthetic data set and the encoder and the segmentation branch was subsequently re-trained on the 144 real images to fine-tune network weights for the segmentation task.
The training was initiated with a learning rate of 1e-5 and ADAM solver was used as the optimizer. The model was trained on the synthetic images for $20$ epochs and on the real images for $10$ epochs.

\section{\hspace{14pt}Results}

The virtual environment has been used to create example mock-up AR-visualizations that have the sole purpose of illustrating the concept of the proposed AR-based suturing support (Fig. \ref{fig:Examples}). In a final application scenario, these visualizations must be connected to the recovered 3D position of the needle and the instrument. We will address this connection in future work. 

The network produces a $3$ channel output for the multi-class segmentation task. The prediction of each object class is obtained in a different channel. Prediction accuracy was evaluated on an independent test set consisting of synthetic and real images. 
The network achieved a dice score of 0.39 for the needle on synthetic images and a much higher dice score of 0.67 on real images. The instruments achieved a dice score of 0.95 (synthetic) and 0.81 (real). 

\begin{figure*}[t]
   \begin{center}
 \includegraphics[width=0.8\linewidth]{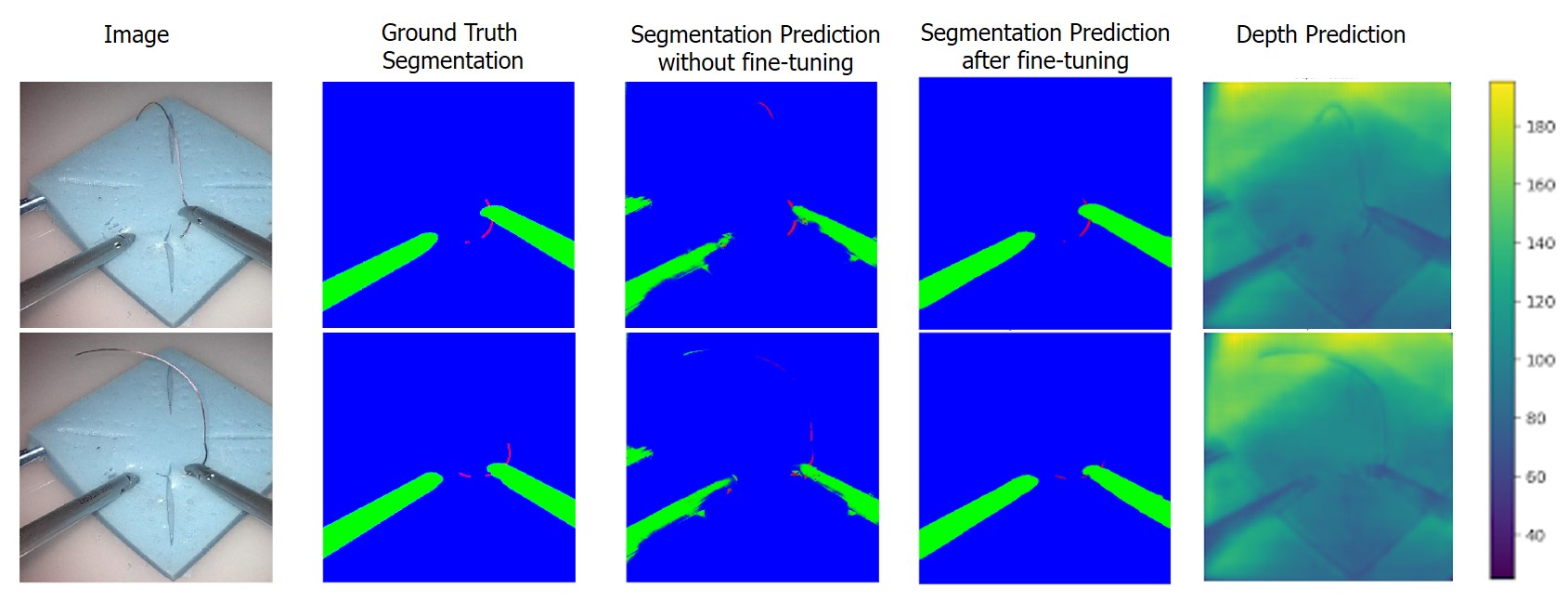}
   \end{center}
   \caption[] 
   { \label{fig:Results1} Examples of network prediction for segmentation and depth estimation on real test data. }
\end{figure*}

Mean absolute error 
was used as the metric for estimating the model's performance in the depth map prediction, with $\hat{y}$ representing the corresponding ground truth information. 
\begin{equation} \label{eq:1}
    MAE = \frac{1}{n}\sum_{j=1}^{n}|y_j - \hat{y_j}|
\end{equation}
The model gave a mean absolute error of 6.5mm on the synthetic test data set.
The dice scores of individual classes is shown in Table \ref{tab:table1}.
Qualitative results of the predictions on real and synthetic images are depicted in Fig. \ref{fig:Results} and \ref{fig:Results1}. The model was able to produce a segmentation and depth estimation of the scene even when encountered with challenges like occlusion.


\begin{table}[h]
  \begin{center}
    \begin{tabular}{l|c|c|c} 
      ~ & \textbf{Synthetic Images} & 
      \textbf{Real Images} &  \textbf{Real Images}\\
      ~ & ~ & (without fine-tuning) & (after fine-tuning) \\
      \hline
      Dice Needle & 0.39 & 0.27 & 0.67 \\
      Dice Instruments & 0.95 & 0.31 & 0.81 \\
      \hline
      MAE depth (mm)& 6.5 & - & - \\
    \end{tabular}
  \end{center}
  \caption{Results of different object classes on real and synthetic test data sets.}
  \label{tab:table1}
\end{table}

\section{\hspace{14pt}Conclusion}

The paper describes a novel use-case for an AR-concept to be potentially useful to guide surgical training. The work solves first steps towards this goal with regard to joint segmentation and depth estimation of the needle and the needle holder. In order to overcome scarcity of labels considering dense depth information, a virtual environment was created and subsequently leveraged for training data generation in a supervised deep learning approach. 


With the emergence of larger resolutions of endoscopic images, i.e.\ full high definition or even 4K, it is now possible to better capture the needle more reliably. The proposed method was able to achieve good segmentation and reasonable depth estimation results on both synthetic and real images (cf. Fig. \ref{fig:Results}). We want to further use this information to teach another network for fine 3D localization on the relevant sub-region, since a much higher accuracy is needed for this application which has not been achieved yet. 

While it is not straightforward to obtain reliable depth information from a surgical training scenario due to the lack of available sensors working on such a small scale with sufficient resolution, we have chosen to design the same scene in a virtual environment to create training data. The drawback of this approach is that we can so far only provide qualitative results on depth estimation of the real scene. 


We exploited the advantage of using a virtual environment, which enables generation of high quality training data, potentially on a large scale in the future. Given the latest developments in domain transfer using generative adversarial networks (GANs), we think that we can achieve comparable results in a more complex training scenario \cite{EngelhardtICVTS} or intraoperative scenario. For instance, Pfeiffer et al. \cite{Pfeifer2019} could show that learned anatomical appearances can be mapped on virtual renderings of the liver using GANs. Similarily, Engelhardt et al.\ \cite{Engelhardt2018,Engelhardt2019} showed the improvement of realistic appearance in minimally-invasive endoscopic surgical training for mitral valve repair. Similar approaches could be taken into consideration here.

The presented approach will enable autonomous learning applications without the need for assistance from teaching personnel. Additionally, it could be extremely relevant for surgical skill assessment \cite{Kowalewski2016DevelopmentAV} 
to quantitatively describe the suturing process in addition to, e.g.\ robotic kinematic data. 


\section*{Acknowledgements}
The Titan Xp GPU card used for this research was donated by the NVIDIA Corporation. We furthermore thank Karl Storz for providing the 3D-surface of the needle holder. 

\bibliography{literatur} 
\bibliographystyle{spiebib} 


\end{document}